\documentclass[twocolumn,english,superscriptaddress,nofootinbib,preprintnumbers, aps,longbibliography]{revtex4-1}
\usepackage{babel}
\usepackage{amsmath}
\usepackage{graphicx}
\usepackage[caption=false]{subfig}
\usepackage{amssymb}
\usepackage{soul}
\usepackage{tensor}
\usepackage{float}
\usepackage{subfig}
\usepackage{hyperref}

\usepackage{xcolor}
\definecolor{cblue}{RGB}{16,78,139}
\definecolor{cred}{RGB}{139,37,0}
\definecolor{cgreen}{RGB}{0,139,0}

\usepackage[left=2cm, right=2cm, top=2cm, bottom=2cm]{geometry}



\begin{document}


\title{Modifications to Gravitational Wave Equation from Canonical Quantum Gravity}

\author{Andrea Dapor}
	\email{adapor1@lsu.edu} 
\affiliation{Department of Physics and Astronomy, Louisiana State University, Baton Rouge, LA 70803, USA}

\author{Klaus Liegener}
\email{klaus.liegener@desy.de}
\affiliation{II. Institute for Theoretical Physics, University of Hamburg, Luruper Chausee 149, 22761 Hamburg, Germany}

\date{\today{}}

\begin{abstract}
It is expected that the quantum nature of spacetime leaves its imprint in all semiclassical gravitational systems, at least in certain regimes, including gravitational waves. In this paper we investigate such imprints on gravitational waves within a specific framework: space is assumed to be discrete (in the form of a regular cubic lattice), and this discrete geometry is quantised following Dirac's canonical quantisation scheme. The semiclassical behavior is then extracted by promoting the expectation value of the Hamiltonian operator on a semiclassical state to an effective Hamiltonian. Considering a family of semiclassical states representing small tensor perturbations to Minkowski background, we derive a quantum-corrected effective wave equation. The deviations from the classical gravitational wave equation are found to be encoded in a modified dispersion relation and controlled by the discreteness parameter of the underlying lattice. For finite discretisations, several interesting effects appear: we investigate the thermodynamical properties of these modified gravitons and, under certain assumptions, derive the tensor power spectrum of the cosmic microwave background. The latter is found to deviate from the classical prediction, in that an amplification of UV modes takes place. We discuss under what circumstances such effect can be in agreement with observations.

\end{abstract}

\maketitle
%

\section{Introduction}
The observation  of gravitational waves (GW) in recent years has opened a new window for insights into the cosmos \cite{Hughes:2014yia,Abbott:2016blz,TheLIGOScientific:2016src,TheLIGOScientific:2017qsa}. This manifested itself in the emergence of the era of multi-messenger astronomy on the one hand \cite{Branchesi:2016vef,MMA17} and on the other hand with the further search for gravitational waves of primordial origin \cite{Krauss:2010ty}. Such waves might soon be detectable, and are expected to lead to a new understanding of the early universe. This includes the possibility to probe the so far unknown frontier of quantum gravity. Indeed, first approaches exist which use GW to narrow down the possibility of {\it discrete spacetimes} \cite{AmelinoCamelia:1998ax} featuring an intrinsic discreteness scale typically associated to the Planck length. However before these phenomenological tests can be carried out fully, it is important to understand the theoretical predictions about the propagation of GW from a given theory of quantum gravity.

The present work will elucidate on this endeavour. We will focus on a particular approach towards quantum gravity, namely {\it Loop Quantum Gravity} (LQG) \cite{Gambini:1996ik,Rovelli:2004tv,Ashtekar:2004eh,Thi08}. This candidate for canonical quantum gravity has matured in the last decades towards a mathematically well-defined theory. It is a quantum field theory on a continuous manifold, however the observables of geometrical quantities (such as areas and volumes)  come {\it naturally} with discrete spectra \cite{Ashtekar:1986yd,Rovelli:1994ge,Ashtekar:1996eg,Ashtekar:1997fb}. Whence, one could expect discrete features of quantum geometry to emerge at the level of observations. Including this discreteness into reduced models led to many interesting concepts: for example, Loop Quantum Cosmology (a quantisation of the cosmological sector of General Relativity using techniques from LQG) resolved the initial singularity via a smooth bounce \cite{Boj08,APS06a,APS06c,Assanioussi:2018hee}. However, the evolution of such cosmological models was prone to discretisation ambiguities. Only recent studies enabled to draw a connection between the reduced dynamics and the Hamiltonian constraint of the full theory \cite{DL17}. This was achieved via coherent states sharply peaked on a discrete geometry \cite{Thiemann:2002vj,Thiemann:2000bw}: interpreting the expectation value of the Hamiltonian constraint operator thereon as an effective Hamiltonian extracts the quantum corrections due to the underlying discrete structure \cite{Thiemann:2000ca,Thiemann:2000bx,Giesel:2006uk,Alesci:2014uha,DL17b} in the form of a modified {\it effective dynamics} for the canonical variables. This paper applies this framework to the case of GW propagating on flat Minkowski background.

Section \ref{s2_classic} of this paper revisits the classical formulation of GW in the Hamiltonian description. Then, we introduce in section \ref{s3_LQG} the necessary formalism of LQG and how coherent states for GW on Minkowski spacetime are constructed. A necessary condition is the perturbative treatment of the modes, as their smallness is necessary to obey the linearised Einstein equations. We will outline how this simplifies the formalism and what result is found for the expectation value of the Hamiltonian constraint, which is then thought of as an effective Hamiltonian on the phase space of linearised gravity. Said effective constraint differs from its classical continuous counterpart, the modifications being controlled by the discreteness parameter $\epsilon$ stemming from the discretised quantum geometry. In section \ref{s4_Dispersion relation}, the modified Hamilton's equation for GW are derived and analytically solved, leading to waves with a specific dispersion relation. In section \ref{s5_ Graviton}, the modified dispersion relation for GW in LQG is studied. The propagating modes can be described as gravitons, for which a standard Fock quantisation is possible. We will describe their thermodynamical properties. Section \ref{s6_cosmology} presents a toy model to investigate whether the modified dispersion relation can lead to observable modifications to the Cosmic Microwave Background (CMB) tensor power spectrum. Finally, section \ref{s7_conclusion} concludes with a comparison to existing literature and an outlook of further research directions.

\section{Canonical description}
\label{s2_classic}
On manifold $\mathcal{M}=\sigma \times \mathbb{R}$ (with $\sigma \cong [0,L]^3$ a torus with periodic boundary conditions) 
 we consider for the spacetime metric a perturbation around flat Minkowski background:
\begin{align}\label{metric}
g_{\mu\nu}=\eta_{\mu\nu}+h_{\mu\nu}
\end{align}
where $h$ is sufficiently small so that its quadratic orders can be neglected. 
By adopting the {\it transverse traceless gauge}, its only non vanishing elements are
\begin{align}
h_{xx}=-h_{yy}=h_+,\;\;\;\;\;\;\; h_{xy}=h_{yx}= h_\times
\end{align}
which depend only on direction $z$ and time $t$.

In the Hamiltonian formulation of gravity, Einstein's equations become equivalent to a totally constrained system \cite{Dirac:1958sc,Arnowitt:1962hi}. Expressed in the Ashtekar-Barbero variables $(E^a_I,A_a^I)$ \cite{Ashtekar:1986yd,BarberoG.:1993aa}, the constraints are
\begin{align}
\begin{array}{c}
C=\dfrac{\epsilon_{IJK}E^a_JE^b_K}{\kappa \sqrt{\det E}}(F^I_{ab}-(1+\beta^2)\epsilon_{IMN}K^M_aK^N_b)
\\
\\
C_a=\dfrac{2}{\kappa\beta}F_{ab}^JE^b_J,\hspace{25pt}
G_J=\partial_a E^a_J+\epsilon_{JKL}A^K_a E^a_L
\end{array}
\end{align}
where $\kappa=16\pi G$ and $\beta>0$ is the Immirzi parameter. $K$ is the extrinsic curvature and $F$ the gauge curvature of connection $A$. The connection and its canonical conjugated momentum, the triad $E$, can be computed for metric (\ref{metric}) (by neglecting $\mathcal{O}(h^2)$ contributions) and read
\begin{align}
\begin{array}{c}
E^1_1=1-\dfrac{h_+}{2}, \ \  E^2_2=1+\dfrac{h_+}{2}, \ \  E^1_2=E^2_1=-\dfrac{h_{\times}}{2}
\\
\\
E^3_3=1, \ \ \ A^1_1=A^2_2=-\beta\dfrac{p_+}{2}, \ \ \ A^2_1=A^1_2=\beta\dfrac{p_\times}{2}
\end{array}
\end{align}
where $p_{+/\times}$ is the canonical conjugated momentum to $h_{+/\times}$ respectively, i.e., $\{h_i(z),p_j(z')\}=\kappa\delta^i_j \delta(z,z')/L^2$.

The Hamiltonian of General Relativity is
\begin{align}
H=\int_{\sigma} {\rm d}x{\rm d}y {\rm d}z\; (N C+N^a C_a)
\end{align}
with lapse function $N$ and shift vector $N^a$. We will gauge fix $N=1$ and $N^a=0$ such that for a spacetime given by (\ref{metric}) the Hamiltonian becomes (up to a boundary term):
\begin{align}
H = L^2 \frac{1}{2\kappa}\int {\rm d}z\; [\dot{h}_+^2+(\partial_z h_+)^2+\dot{h}_\times^2+(\partial_z h_\times)^2]
\end{align}
This describes two decoupled, massless free scalar fields in one dimension. The solutions are thus classical waves.

\section{Loop Quantisation}
\label{s3_LQG}
Since we are interested in the consequences for GW from discrete spatial manifolds, we now introduce an ad-hoc discretisation of $\sigma$. This discretisation is a cubic lattice with edges of coordinate length $\epsilon$ (and its dual cell complex); the number of vertices $v$ in each direction is $N=L/\epsilon$. Keeping $\epsilon$ finite and only considering finitely many degrees of freedom described by smearings along the edges and faces of the lattice allows to proceed with the canonical quantisation of General Relativity in a well-defined manner, analogously to the quantisation procedure of LQG. We emphasize that discretisation of space is not a necessity in LQG, so we regard it as an additional postulate.

The variables we are interested in are ${\rm SU}(2)$-valued holonomies of $A$ along the edges $e$ of the lattice and gauge-covariant fluxes\footnote
{
With the choice in \cite{ThiVII_00,SL19b}, these coincide with conventional fluxes up to $\mathcal{O}(h^2)$.
}
of $E$ across the dual faces $S_e$ for each edge $e$:
\begin{align}
h(e)= \mathcal{P}\exp\left(-\int_e A\right),\hspace{8pt}  P(e)=\int_{S_e} * E+\mathcal{O}(h^2)
\end{align}
where $A=A^I\tau_I,\;E=E_I\tau_I$ and $\tau_I=-i \sigma_I/2$, with $\sigma_I$ the Pauli matrices.

For a lattice whose edges are oriented along the coordinate directions, the discretisation of (\ref{metric}) gives (neglecting $\mathcal{O}(h^2)$-contributions)
\begin{align}\label{discGeo}
h(e_1)&={\rm id}-\epsilon\beta(p_+ \tau_1+p_\times \tau_2)/2\\
h(e_2)&={\rm id}-\epsilon\beta(p_\times \tau_1-p_+ \tau_2)/2,\hspace{10pt} h(e_3)={\rm id}\nonumber\\
P(e_1)&=\epsilon^2\tau_1(1- B_+(z))-\epsilon^2\tau_2 B_\times(z)\\
P(e_2)&=\epsilon^2\tau_2(1+ B_+(z))-\epsilon^2\tau_1 B_\times(z),\hspace{10pt} P(e_3)=\epsilon^2\tau_3\nonumber
\end{align}
with 
\begin{align} \label{def-of-Bs}
B_i(z):=\frac{1}{2\epsilon}\int_{z-\epsilon/2}^{z+\epsilon/2}{\rm d}u\; h_i(u)
\end{align}
These quantities describe the discrete spatial geometry on the initial-time hypersurface.

Canonical quantisation can now be performed, leading to the Hilbert space of square-integrable functions over ${\rm SU(2)}$ on each edge $e$: $\mathcal{H}_e=L_2({\rm SU}(2),d\mu_H)$ with $\mu_H$ the Haar measure. A coherent state $\Psi\in\mathcal{H}=\otimes_e \mathcal{H}_e$ for the discretised geometry (\ref{discGeo}) of GW can now be explicitly constructed following \cite{Thiemann:2002vj,Thiemann:2000bw,Thiemann:2000ca,Thiemann:2000bx}:
\begin{align}
\Psi&= \prod_{e} \psi_{e} \\
\psi_{e}(g)&=\sum_j (2j+1)e^{-j(j+1)t/2}{\rm Tr}^{(j)}(H^\dagger_{e} g)\\
H_{e}&=\exp[-it\;P(e)/(\hbar\kappa\beta)]h(e)
\end{align}
Here, $t>0$ describes the spread of the state: $\Psi$ is sharply peaked, that is,
\begin{align}
\begin{array}{c}
\langle \Psi, \hat{h}(e_i) \Psi \rangle = h(e_i) [1+\mathcal{O}(t)]
\\
\\
\langle \Psi, \hat{P}(e_i) \Psi \rangle = P(e_i)[1+\mathcal{O}(t)]
\end{array}
\end{align}
where $\hat{h}$ is the multiplication operator and $\hat{P}$ the right-invariant vector field.\footnote
{
$t$ is in principle a free parameter in $\Psi$. However, we can adopt $t=\|h\|^2$, which implies that all corrections of non-zero order in $t$ can be neglected in the linearisation.
}

This peakedness also extends to more complicated observables built from these basic operators, leading to the realisation that the expectation value of any quantity on the quantum state will result in the corresponding classical discretisation (up to quantum corrections proportional to the spread of the state). This leads to the conjecture that the dynamical evolution at the quantum level is well approximated by the dynamics generated by the discretised Hamiltonian of the system, that is, the leading order of the expectation value of $\hat H = \sum_v \hat C(v)$. In its most prominent quantisation \cite{Thiemann:1996aw,Thiemann:1996av}, operator $\hat C(v)$ reads
\begin{align}
\hat{C}(v)=\hat{C}_E(v) +\hat{C}_L(v)
\end{align}
with
\begin{widetext}
\begin{align}
&\hat{C}_L(v)={\frac{4(1+\beta^2)}{\kappa^4\beta^7i\hbar^5}}\sum_{e\cap e'\cap e''=v}\epsilon(e,e'e,''){\rm Tr}^{(1/2)}(\hat h(e)\left[\hat h (e)^\dagger,\hat K\right]\hat h(e')\left[\hat h (e')^\dagger,\hat K\right]\hat h(e'')\left[\hat h (e'')^\dagger,\hat V\right])\\
&\hat{C}_E(v)={\frac{-1}{12\kappa^2\beta i\hbar}}\sum_{e\cap e'\cap e''=v}\epsilon(e,e',e''){\rm Tr^{(1/2)}}((\hat h (\Box_{ee'})-\hat h^\dagger(\Box_{ee'}))\hat h(e'')\left[\hat h(e'')^\dagger,\hat V\right]),\hspace{40pt}\hat K= \sum_v\left[\hat C_E(v),\hat V\right]\nonumber
\end{align}
where $\Box_{ee'}$ is the minimal plaquette spanned by $e$ and $e'$, while $\hat V$ is the volume of the whole spatial manifold $\sigma$.

Using the fact, that the expectation value agrees at leading order in $t$ with its classical regularised expression, one can perform (after a lengthy computation) the reduction to the phase space spanned by $h_+(z), h_\times(z), p_+(z), p_\times(z)$.
The expectation value at each vertex $v$ is found to be (we define $f^{\pm m}:=f(z \pm m\epsilon)$ for any function $f$)
\begin{align}\label{ExpValue}
\langle\Psi,\hat{C}_E(v)\Psi\rangle&=
-\frac{\epsilon^3\beta^2}{2\kappa}\left[p_+^2+p_\times^2+\frac{2B_+}{\beta}\frac{p_\times^+-p_\times^-}{\epsilon}-\frac{2B_\times}{\beta}\frac{p_+^+-p^-_+}{\epsilon}\right]
+\mathcal{O}(t)\\
\langle\Psi,\hat{C}_L(v)\Psi\rangle&=
\epsilon^3\frac{1+\beta^2}{2\kappa}\left[\left(\frac{B_+^+-B^-_+}{\beta\epsilon}-\frac{p_\times^++2p_\times+p^-_\times}{4}\right)^2+
\left(\frac{B_\times^+-B^-_\times}{\beta\epsilon}+\frac{p_+^++2p_++p^-_+}{4}\right)^2
\right]
+\mathcal{O}(t)\nonumber
\end{align}
\end{widetext}
where, due to homogeneity in $x$ and $y$ directions, the quantities on the rhs depend only on the $z$-coordinate of $v$ (i.e., $z \in \{n\epsilon : n = 0,\pm 1,.., \pm N/2\}$).

\section{Modified dispersion relation}
\label{s4_Dispersion relation}
The non-trivial modifications (\ref{ExpValue}) of the Hamiltonian for GW due to the discreteness of space, can now be used to extract physical predictions upon adopting the conjecture mentioned above: the dynamics on the discrete phase space of General Relativity for initial data belonging to a certain symmetry class (i.e., the reduced phase space of GW in our case) can be described completely on the reduced phase space by using the reduced, discrete Hamiltonian as generator of time evolution.

From now on, we denote $p_+=p_1,p_\times=p_2$ (and similar for $h$ with $1,2\in\mathbb{Z}_2$) and work with the above conjecture. Then, from
\begin{align}
H_{\rm eff}=H_{\rm eff}(h_i,p_i):= \langle\Psi, \sum_{v}\hat C(v) \Psi\rangle
\end{align}
we can derive the Hamilton equations on the reduced, discrete phase space (using $\dot{f}=\{f, H_{\rm eff}\}$ for any observable $f=f(h_i,p_i)$):

\begin{align}\label{Hamiltonequation-p}
\dot{p}_i&=(-)^{i+1}\frac{\beta}{2}\frac{p_{i+1}^+-p_{i+1}^-}{\epsilon}+\\
&+\frac{1+\beta^2}{2\beta\epsilon}\left(\epsilon\mathbb{B}_i+(-)^i\frac{p^{+2}_{i+1}+2(p^+_{i+1}-p^-_{i+1})-p^{-2}_{i+1}}{4}
\right)\nonumber
\end{align}

\begin{align}\label{Hamiltonequation-B}
\dot{\mathbb{B}}_i&=-\frac{\beta}{2}\left(\frac{p^{+2}_i-2p_i+p^{-2}_i}{\epsilon^2}-(-)^i\frac{\mathbb{B}^+_{i+1}-\mathbb{B}^-_{i+1}}{\epsilon}\right)\\
&+\frac{1+\beta^2}{8\beta}\left(\Delta_i-(-)^{i}\frac{\mathbb{B}^{+2}_{i+1}+2(\mathbb{B}^+_{i+1}-\mathbb{B}^-_{i+1})-\mathbb{B}^{-2}_{i+1}}{\epsilon}\right)\nonumber
\end{align}
with $\mathbb{B}_i := (B_i^{+2}+B_i^{-2}) /(\epsilon^2\beta)$ and
\begin{align}
\Delta_i:=&\frac{1}{4\epsilon^2}[p^{+4}_i+4p^{+3}_i+4p^{+2}_i-4p_i^{+1}-10p_i\nonumber\\
&-4p_i^-+4p^{-2}_i+4p^{-3}_i+p_i^{-4}]
\end{align}
Equation (\ref{Hamiltonequation-p}) can be inverted for $\mathbb{B}$ in order to express the right hand side of (\ref{Hamiltonequation-B}) as function of $p_i,\dot{p}_i$ only, i.e., $\dot{\mathbb{B}}_i=f(p_+,p_\times,\dot{p}_+,\dot{p}_\times)$. This expression can be used when taking the time derivative on equation (\ref{Hamiltonequation-p}) to obtain
\begin{align}\label{EoMofMomenta}
\ddot{p}_i=\frac{\beta^2}{8\epsilon^2}(sp_i^{+3}-2p^{+2}_i-sp_i^++4p_i-sp_i^--2p^{-2}_i+sp^{-3}_i)
\end{align}
with $s=(1+\beta^2)/\beta^2$. Note that the two degrees of freedom corresponding to $i=1$ and $i=2$ decouple.

To solve (\ref{EoMofMomenta}), we first extend the size of the box to infinity by keeping the lattice spacing $\epsilon$ constant: $L \to \infty$ and $N \to \infty$ such that $L/N = \epsilon$ constant. Then, the {\it ansatz}
\begin{align} \label{p-sol-temp}
p(z) = \dfrac{\epsilon}{2\pi} \int_{\mathcal{B}} {\rm d}k \ e^{ikz} \ u(k)
\end{align}
with $k$ taking values in the {\it Brillouin zone} $\mathcal{B} := [-\pi/\epsilon, \pi/\epsilon]$, solves (\ref{EoMofMomenta}) if the Fourier coefficients $u(k)$ obey the equation
\begin{align}\label{New_waveq}
\ddot{u}(k)=-\omega(k)^2\; u(k)
\end{align}
with
\begin{align}
\omega(k)^2=\frac{\sin(k\epsilon)^2}{\epsilon^2}((1+\beta^2)\cos(k\epsilon)-\beta^2)\label{New_Dispersion_Relation}
\end{align}
Observe that oscillatory modes are only possible for $\omega(k)^2>0$ (modes out of this range do not propagate), therefore we restrict our attention to $|k|< k_o$ with
\begin{align} \label{max-q}
k_o:=\frac{1}{\epsilon}\arccos\left(\frac{\beta^2}{1+\beta^2}\right) < \dfrac{\pi}{\epsilon}
\end{align}
For such modes, the solution to (\ref{New_waveq}) is simply
\begin{align}\label{New-wavsol}
u(k,t) = C_k^+ e^{i\omega(k) t} + C_k^- e^{-i\omega(k) t}
\end{align}
with $C_k^\pm$ complex constants. At this point, we plug (\ref{p-sol-temp}) (with $u(k)$ given by (\ref{New-wavsol})) in (\ref{Hamiltonequation-p}) and solve for $\mathbb{B}_{i}(z,t)$: this finally leads to expression
\begin{align}\label{free-wave}
B_{i}(z,t) = \dfrac{\epsilon}{2\pi} \int_{-k_o}^{+k_o} {\rm d}k \ [a_{i,k} e^{i(kz+\omega(k)t)}+a^*_{i,-k}e^{i(kz-\omega(k)t)}]
\end{align}
The coefficients $a_{i,k} \in \mathbb C$ are related to $C_k^\pm$ by complicated expressions but, since $C_k^\pm$ are anyway generic and since $B_i$ are the quantities most closely related to the metric components $h_1$ and $h_2$ (see equation (\ref{def-of-Bs})), we take $a_{i,k}$ to be the fundamental quantities that characterise the field $B_i$.

Now, although $B_i$ is essentially a linear combination of plane waves, such waves have a modified dispersion relation, in contrast to the classical $\omega_{\rm cl}(k)^2=k^2$. This modification appears due to the discrete structure of space, which is controlled by the lattice parameter $\epsilon$. Indeed, $\lim_{\epsilon\to 0}\omega = \omega_{\rm cl}$, hence one can expect the modifications due to $\omega(k)$ to be comparably small and hard to detect for extremely fine discretisations. Conversely, for gravitational waves with high momentum $k$ there might be a measurable departure from the classical predictions. We will discuss some potentially measurable effects in sections \ref{s5_ Graviton} and \ref{s6_cosmology}.

Let us close this section with a remark: observe that ``classical'' waves have wavelengths much larger than the discreteness scale $\epsilon$, which means that $k\epsilon \ll 1$, whence classical waves move at the speed of light. However, the general formula for the speed of waves reads
\begin{align}
v_k:=\frac{d|\omega(k)|}{dk}|_{k=0}=1-\frac{5+3\beta^2}{4}k^2\epsilon^2+\mathcal{O}(k^4\epsilon^4)
\end{align}
revealing that gravitons with short wavelengths propagate {\it slower} than light.

\section{Effective Graviton}
\label{s5_ Graviton}
A surprising aspect of the analysis in the previous section was that the only modifications of GW due to the presence of discrete spacetime is a modified dispersion relation with $\omega(k)$ given in (\ref{New_Dispersion_Relation}). In this section we will study the consequences of such a modified dispersion relation for gravitons, understood as a Fock quantisation of the degrees of freedom $B_i$ obtained in the previous section. We emphasize that this ``re-quantisation'' is a model capturing only certain aspects of the original quantum gravity theory (described in section \ref{s3_LQG}).

Since we are dealing with free fields, we can perform the quantisation as in standard quantum field theory, by promoting $a_{i,k}$ of (\ref{free-wave}) to operators on the Fock space $F(\mathcal{H})$:
\begin{align}
\begin{array}{c}
\tensor{a}{_{i,k}},a^*_{i,k}\mapsto \tensor{\hat{a}}{_{i,k}},\hat a_{i,k}^\dagger,\;\; [\tensor{\hat a}{_{i,k}},\hat a^\dagger_{i'k'}]=\delta_{i,i'}\delta(k,k') \hat I
\\
\\
\mathcal{H} := \left\{\int {\rm d}k \ \hat a^\dagger_{i,k}f(k) |0\rangle,\;f\in L_2(\mathbb R)\right\}
\end{array}
\end{align}
Operator $\hat a_{i,k}^\dagger$ creates a graviton with polarisation $i$ and momentum $k$. Knowing that the classical system is a set of infinitely  many harmonic oscillators with frequency $\omega(k)$, the Hamiltonian describing the system is\footnote
{
Indeed, since the solutions of (\ref{EoMofMomenta}) are plane waves (when restricting to $|k|<k_o$), there exists a canonical transformation bringing the Hamiltonian into harmonic oscillator form.
}
\begin{align}
\hat H =\hbar \sum_{i=1,2} \int_{-k_o}^{+k_o} {\rm d}k \ \omega(k) \; \hat a^\dagger_{i,k} \tensor{\hat a}{_{i,k}}
\end{align}

\begin{figure}[h]
\includegraphics[width=8cm]{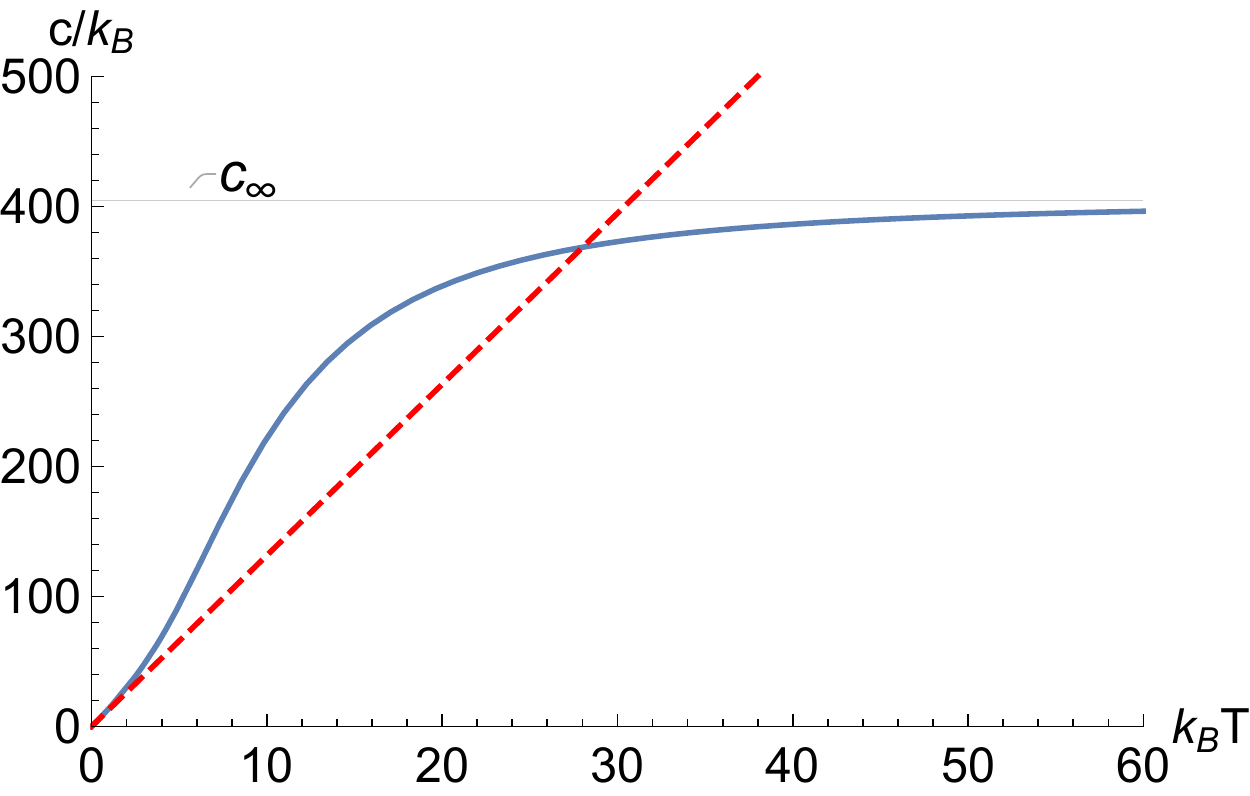}
\caption{Comparison of the specific heat capacity $c(T)$ with $\beta=0.2375$ in the two cases: classical dispersion relation in the continuum (red, dashed), modified dispersion relation for $\epsilon=0.015$ (blue, solid). In contrast to the continuum model, the specific heat capacity of gravitons in the discrete approaches a constant $c_\infty$ for $T\to\infty$.}
\label{fig:1}
\end{figure}

Since we can interpret $\hat a^\dagger_{i,k} \tensor{\hat a}{_{i,k}}$ as the operator $\hat n_{i,k}$ measuring the number of gravitons with polarisation $i$ and momentum $k$, and knowing that for a thermal state at temperature $T$ it is $\langle \hat n_{i,k} \rangle = (e^{\hbar \omega(k)/(k_BT)} - 1)^{-1}$ (where $k_B$ is Boltzmann constant), we find for the total energy
\begin{align}
\bar E := \langle \hat H \rangle = 2 \int_{-k_o}^{+k_o} {\rm d}k \ \dfrac{\hbar \omega(k)}{e^{\hbar \omega(k)/(k_BT)} - 1}
\end{align}
To understand the modifications due to discreteness in the behaviour of energy with respect to temperature, it is instructive to look at the specific heat capacity $c$:
\begin{align}
c &:= \frac{\partial \bar E}{\partial T} =\label{heat_cap}\\
 &= \dfrac{2}{k_B T^2} \int_{-k_o}^{+k_o} {\rm d}k \left(\dfrac{\hbar \omega(k)}{e^{\hbar \omega(k)/(k_BT)} - 1}\right)^2 e^{\hbar \omega(k)/(k_BT)}\nonumber
\end{align}
In the large temperature limit, $c$ approaches the constant $c_\infty=2k_BM$, with $M = 2k_o$ the size of integration. This is a strikingly different behaviour than that of the continuum model. Indeed, for $\epsilon\to0$ it is $k_0\to\infty$, and the integral in (\ref{heat_cap}) can be computed analytically to be $4\pi T /3$. We show in figure \ref{fig:1} how this deviation manifests itself for the case $\epsilon=0.015$ in Planck units (in blue) compared to the continuum case (in dashed red). It transpires that, compared to the usual dispersion relation, the discrete spacetime causes first an increase for the specific heat capacity, but then approaches quickly a constant, leading to a novel behaviour for high temperatures (not unlike what is found for phonons in crystals).

\section{A toy model for primordial tensor modes}
\label{s6_cosmology}
An important application for linearised gravity is the very early universe, where perturbation theory can be used to reproduce the observable power spectrum to high accuracy. However, instead of establishing a consistent quantum theory of perturbations (see\cite{Gomar:2015oea,ElizagaNavascues:2016vqw,Schander:2019wuq} for guidelines), we will assume in this section that the modified dispersion relation that was previously derived for GW on Minkowski spacetime can be directly used for the tensor perturbations in isotropic Friedmann-Lema\^itre-Roberston-Walker cosmology. The only difference is that the general form of the graviton field is not (\ref{free-wave}), but rather (using conformal time $\eta$)
\begin{align} \label{field-deco}
	B_i(z,\eta) \propto \int \dfrac{{\rm d}k}{s(\eta)} \left[\tensor{a}{_{k}} e^{ikz} \xi_k(\eta) + a_{k}^\dag e^{-ikz} \xi_k(\eta)^*\right]
\end{align}
where $s$ is the scale factor, which is here taken to undergo inflation driven by a minimally-coupled free scalar field $\phi$ with mass $m$. In the continuum theory the field $B_i$ satisfies $\Box B_i = 0$, which leads to the equation $\xi_k'' + (k^2 - s''/s) \xi_k = 0$ for the mode function $\xi_k(\eta)$, with $f' := df/d\eta$. It is therefore plausible that the effect of the discrete lattice amounts to the replacement $k \to \omega(k)$. In the following, we thus consider mode functions satisfying
	\begin{align} \label{xi-eq}
	\xi_k'' + \left(\omega(k)^2 - \dfrac{s''}{s}\right) \xi_k = 0
	\end{align}
where $\omega(k)$ is given in (\ref{New_Dispersion_Relation}).

\begin{figure}[h]
		\includegraphics[width=8.5cm]{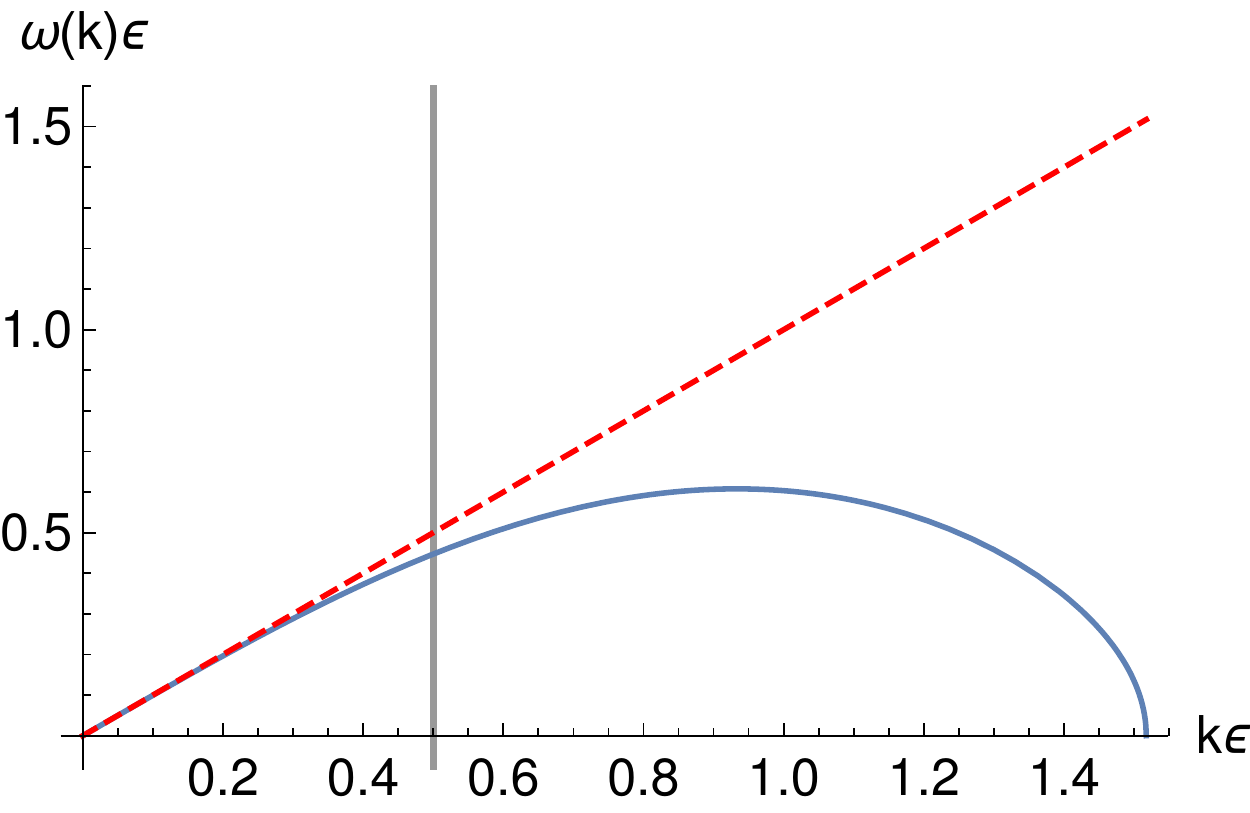}
		\centering
		\caption{Comparison of the two dispersion relations: the classical one (red, dashed) and the modified one (blue, solid) given by (\ref{New_Dispersion_Relation}). The vertical solid line denotes the maximal observable mode $k_{\rm max} = 10^3 k_*$ (with $k_*$ the pivot mode corresponding to $0.002 \ \text{Mpc}^{-1}$ today), under the choices $\epsilon = 0.5/k_{\rm max} = 0.015$ and $\beta=0.2375$ and with inflation driven by a scalar field with mass $m = 1.3 \times 10^{-6}$ and value $\phi(\eta_o) = 3.13$ at the beginning of inflation ($d\phi/dt|_{\eta_o} = 0$) -- all quantities expressed in Planck units.}
		\label{fig:4}
\end{figure}

The space of complex solutions to equation (\ref{xi-eq}) can be parametrised by a single solution $\xi_k$ (and its conjugated $\xi_k^*$), whose choice determines the decomposition into creation/annihilation operators given in (\ref{field-deco}): in other words, choosing a solution of (\ref{xi-eq}), $\xi_k$ (for every $k$), corresponds to choosing a vacuum state. A natural choice is possible for those modes which satisfy the adiabatic condition $\omega(k)^2 \gg s''/s$, since in that case the equation reduces to the one found in Minkowski spacetime, for which Poincar\'e vacuum is uniquely defined. Now, since $\omega(k)$ is constant in time while $s''/s$ grows during inflation, it is possible to satisfy the adiabatic condition for any given mode $k$ by going sufficiently far in the past. How far is enough? To find out, consider the observable window in the CMB, $[k_{\rm min}, k_{\rm max}] = k_* [10^{-1},10^3]$ (with $k_*$ the pivot mode whose physical wavenumber $k_*^P(\eta) := k_*/s(\eta)$ satisfies $k^P_*(\eta_{\rm today}) = 0.002 \ \text{Mpc}^{-1}$), and note that $\omega(k) = k + \mathcal O(k\epsilon)$.\footnote
{
The quantity $k \epsilon$ is invariant under rescaling $s \to \alpha s$, since it is $k \epsilon = k^P(\eta) s(\eta) \epsilon = k^P(\eta) \epsilon^P(\eta)$, where $\epsilon^P(\eta) := \int {\rm d}z \ s(\eta)$ is the physical length of a lattice edge.
} Then, we have three cases:
\begin{enumerate}
\item $\epsilon \approx 0.5/k_{\rm max}$, if we want modifications to the classical power spectrum to fall in the high-$k$ end of the observable range (see figure \ref{fig:4}).
\item $\epsilon \ll 0.5/k_{\rm max}$, if we want no observable modifications to the classical power spectrum.
\item $\epsilon \gg 0.5/k_{\rm max}$, if we want modifications to affect the whole observable range.
\end{enumerate}
Let us focus on the first case, where one can show that $\omega(k)^2 \gg s''(\eta_o)/s(\eta_o)$ for all $k \in [k_{\rm min}, k_{\rm max}]$ as long as $\eta_o$ corresponds to at least $N \approx 62$ e-folds inflation. As a consequence, we can choose for $\xi_k$ the unique solution of (\ref{xi-eq}) with ``Poincar\'e initial conditions''
\begin{align} \label{poincare-ic}
\xi_k(\eta_o) = \dfrac{e^{i \omega_k \eta_o}}{\sqrt{2\omega_k}}, \ \ \ \ \ \xi_k'(\eta_o) = i \omega_k \xi_k(\eta_o)
\end{align}
Given the solution $\xi_k(\eta)$, the tensor power spectrum corresponds to the 2-point-function of $\hat B_i$ evaluated at the end of inflation:
\begin{align}
P_{\mathcal T}(k,\eta) = \dfrac{2\kappa k^3}{\pi^2} \dfrac{|\xi_k(\eta)|^2}{s(\eta)^2}
\end{align}
Note that, soon after a mode $k$ exits the horizon (i.e., satisfies $\omega(k)^2 = s''/s$), its equation (\ref{xi-eq}) reduces to $\xi_k'' - (s''/s) \xi_k = 0$, whose solution is $\xi_k \sim s$. As a consequence, the combination $\xi_k/s$ becomes constant (the mode ``freezes''), and so $P_{\mathcal T}$ for that specific mode will become $\eta$-independent. This means that we do not need to wait until the end of inflation in order to evaluate the power spectrum, but it suffices to take $\eta$ late enough that all modes we are interested in exited the horizon.

\begin{figure}[h]
	\includegraphics[width=8.5cm]{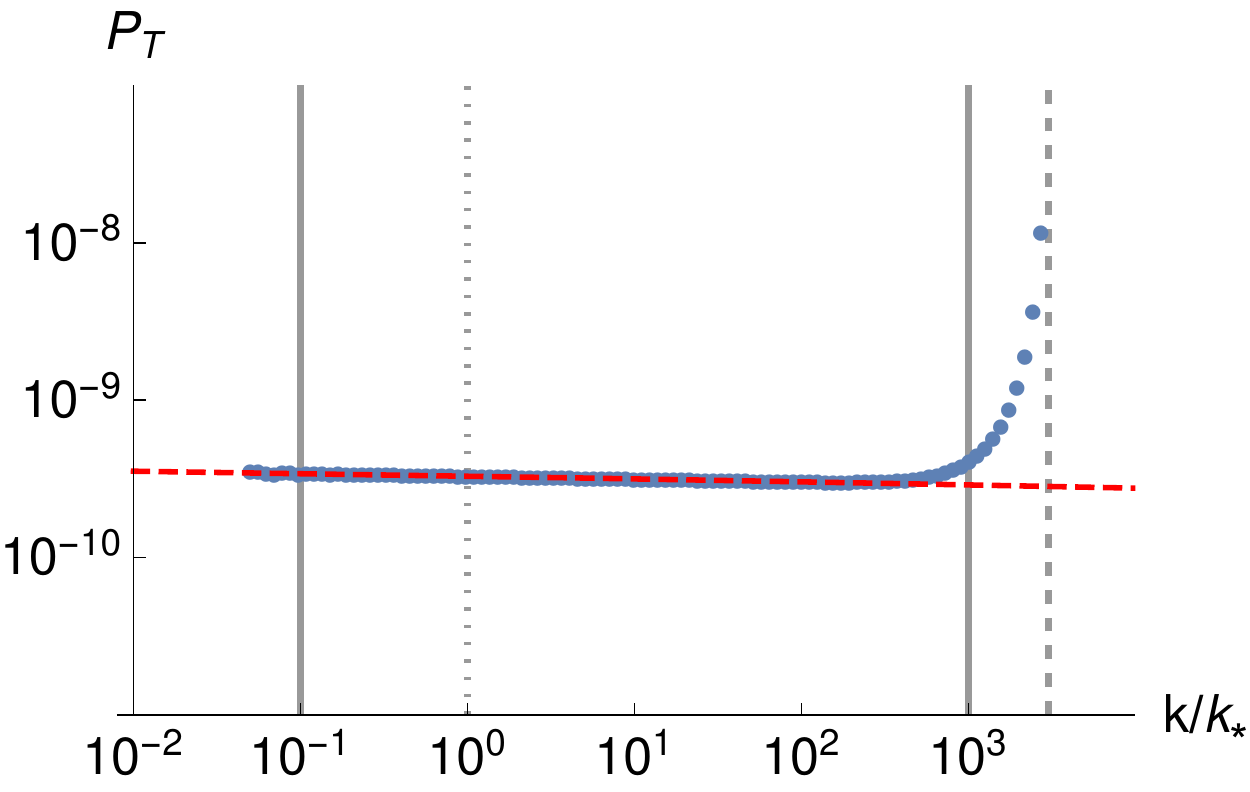}
	\centering
	\caption{Comparison of the tensor power spectrum $P_{\mathcal T}$: the classical prediction obtained from $\omega_{\rm cl}(k) = k$ (red dashed curve) and the prediction obtained from modified dispersion relation $\omega(k)$ (blue dots). In both cases, the mode equation has been solved numerically with initial conditions (\ref{poincare-ic}) at $\eta_o$ corresponding to $N = 62.45$ e-folds, which is achieved by a scalar field with mass $m = 1.3 \times 10^{-6}$ and value $\phi(\eta_o) = 3.13$ at the beginning of inflation ($d\phi/dt|_{\eta_o} = 0$) -- all quantities expressed in Planck units. For the modified case, the choices $\epsilon = 0.015$ and $\beta = 0.2375$ have been made. The vertical solid lines denote the observable region $k \in k_*[10^{-1},10^3]$ (with $k_*$ the pivot mode corresponding to $0.002 \ \text{Mpc}^{-1}$ today), while the dashed one is the maximum propagating mode $k_o$, according to equation (\ref{max-q}).}
	\label{fig:6}
\end{figure}

Since classically it is $\omega_{\rm cl}(k) = k$, modes with higher $k$ exit the horizon at slightly later times: the classical prediction is therefore a power spectrum which is almost $k$-independent (or ``almost scale-invariant'') with a slight tilt downwards, i.e., decreasing from $+\infty$ (for $k=0$) to $-\infty$ (for $k \to \infty$). The expectation for the modified $\omega(k)$ is completely different: $\omega(k)$ grows approximately linearly in $k$ only up to some point, but then it reaches a maximum and goes back down to $0$ (for $k = k_o$); this means that the power spectrum for modes obeying the modified dispersion relation will be almost $k$-independent (with a tilt downwards) up to a minimum value, after which the behaviour turns around and the power spectrum grows (in a non-symmetric fashion), reaching infinity in correspondence of the highest propagating mode $k_o$. This behaviour, and in particular the high-$k$ amplification, is confirmed by a numerical simulation of the system (see figure \ref{fig:6}).

Let us comment on the possible number of e-folds. As mentioned, there is a lower bound $N \gtrsim 62$, which is needed to ensure that the adiabatic condition is satisfied at $\eta_o$ for all observable modes. Interestingly, there might also be a higher bound, depending on the assumptions that one wants to make. On the one hand, if one does not want the amplification to fall in the observable range (i.e., one wants to avoid strong deviations from the classical prediction), one must require $\epsilon^P(\eta) k_{\rm max}^P(\eta) = \epsilon k_{\rm max} \lesssim 0.5$ for all $\eta$, where $k^P(\eta) = k/s(\eta)$ and $\epsilon^P(\eta) = \int {\rm d}z \ s(\eta) = \epsilon s(\eta)$ is the physical length of a lattice edge. On the other hand, one could argue that the minimum physical length in a quantum gravity theory should be something of Planck order, $\ell_{\rm P}$; as $s$ is smallest at the beginning of inflation (in a conservative model, without modifications to the pre-inflationary dynamics), one could therefore require that $\epsilon^P(\eta_o) \gtrsim \ell_{\rm P}$. Putting the two requirements together, one finds that $k_{\rm max}^P(\eta_o) \lesssim 0.5/\ell_{\rm P}$. This puts an upper bound on $N$, since a mode $k^P(\eta_o)$ becomes more and more UV the longer the inflation. In fact, for $N = 62.45$, we find $k^P_{\rm max} \approx 0.95/\ell_{\rm P}$. Thus, if one believes that $\ell_{\rm P}$ is the minimum possible length and does not want too strong a departure from the classical power spectrum, then one concludes that the number of e-folds is very strictly bounded around $N = 62$. Of course, we emphasize that this conclusion is reached within the current toy model, and thus its validity needs to be checked in a more complete theory. Nevertheless, since it only relies on the existence of a minimum length, the general argument could remain valid independently of the details of the theory.

\section{Conclusion}
\label{s7_conclusion}
In this work we considered the effects on linearised gravity due to a discretisation of space. Our approach is based on the framework of Loop Quantum Gravity, a theory which intrinsically features the discreteness of geometrical quantities such as areas. In the Hilbert space of LQG, we chose a family of semiclassical states representing a discrete spatial manifold on which the metric degrees of freedom are sharply peaked on linearised gravity, i.e., gravitational waves on Minkowski background. Taking these states as describing the system at a given time, we computed the expectation value of the Hamiltonian operator, which generates the dynamics in LQG. It was found that such expectation value does not agree with the classical Hamiltonian of gravitational waves but, when used as an effective Hamiltonian on the phase space of linearised gravity, it produces a wave equation featuring a {\it modified dispersion relation}. The modification with respect to the classical wave equation captures the LQG effects due to a discrete spatial manifold.

Let us take a moment to compare our approach to different works in LQG on gravitational waves.\footnote
{
Several works not employing the Ashtekar formalism exist as well, see e.g. \cite{Hoehn:2014qxa}.
}
Since we are working with states in the Hilbert space of full LQG, we differ from early approaches, where it was attempted to quantise only the linearised field theory \cite{Ashtekar:1991mz, Varadarajan:2002ht, Freidel:2003pu}. We also differ from approaches where, similar to LQC, symmetries are implemented prior to quantisation \cite{Hinterleitner:2011rb,Hinterleitner:2017ard, Neville:2013wba,Neville:2013xba}. While these approaches take advantage of a simpler computational setting, it is as of today not established how the quantisation of a reduced theory is connected to the quantisation of the corresponding full theory; therefore, we refrained from taking this route. Instead, working in the full theory required us to perform a lengthy computation, at the end of which the quantum gravity modifications of the Hamiltonian constraint (and their influence for gravitational waves propagation) could be extracted: this is in constrast with works such as \cite{Bojowald:2007cd}, where the modifications are postulated. Of course, the modified dispersion relation we obtained is not necessarily a general feature of LQG: it was obtained from a certain regularisation of the Hamiltonian constraint (see \cite{Bahr:2010cq} for steps towards estimating the influence of renormalisation for linearised gravity), and using a specific family of semiclassical coherent states as initial data (for corrections to linearised gravity in the path integral framework obtained  see \cite{Mikovic:2011fr,Han:2018fmu}). However, the complexifier coherent states used in this paper satisfy many useful properties, enabling analytic calculations and producing a surprisingly simple result.

After deriving the modified dispersion relation, we studied some interesting effects it induces in physically realistic settings. In section \ref{s5_ Graviton} we re-quantised the gravitational waves, thus obtaining ``gravitons'' which include the corrections due to the discreteness of space, and then studied the thermodynamical properties of such particles, finding that they are somehow close to those of phonons in crystals. In section \ref{s6_cosmology} we considered the case where the background is given by a cosmological inflationary spacetime rather than Minkowski: this was done by using the previously derived dispersion relation in the wave equation for the tensor modes. While this is a simplification (which is why we called it a ``toy model''), it allowed us to get an idea of potentially interesting effects in the CMB tensor power spectrum.\footnote
{
We should mention that this toy model differs from other approaches to couple Loop Quantum Cosmology with gravitational waves (see \cite{Mielczarek:2010bh,Sa:2011rm,Agullo:2015tca} and references therein): in those cases, one treats the pertrubations as in standard cosmology, limiting the quantum gravity corrections to the dynamics of the background (usually relevant only in the pre-inflationary phase). In our case, instead, we started our analysis at the beginning of inflation (choosing the vacuum state there), and so we can safely disregard any effect of quantum gravity corrections to the background.
}
In particular, we obtained an amplification in UV modes, which may or may not fall in the observable range depending on the lattice spacing $\epsilon$ and the number of e-folds $N$. The latter was found to be constrained around $N = 62$ if one wants to avoid large departures from the (classically predicted) almost scale-invariance {\it and} if the minimum physical length of a lattice edge is of order $\ell_{\rm P}$. However, we stress that these predictions are obtained from a toy model: to confirm them, a full quantum treatment of background, perturbations and matter \cite{Hossain:2009vd} (which would also allow to calculate the corrections to the scalar power spectrum) has to be developed, possibly on the lines of what was done in this paper for the case of Minkowski background.\\

\noindent\textbf{Acknowledgements:} The authors thank Ivan Agullo and Dimitrios Kranas for many discussions about the tensor power spectrum. They are also grateful to Deepak Vaid, for suggesting this research topic at the 2018 Tux workshop on Quantum Gravity. A.D. acknowledges the support of the NSF Grant No. PHY-1603630. K.L. acknowledges support by the German Research Foundation (DFG) under Germany’s
Excellence Strategy – EXC 2121 ``Quantum Universe'' – 390833306. This work was partially funded by DFG-project BA 4966/1-2.

\twocolumngrid
\bibliography{References}{}
\end{document}